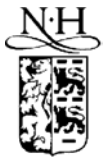

# Radial position of single-site gamma-ray interactions from a parametric pulse shape analysis of germanium detector signals

John L. Orrell,[*] Craig E. Aalseth, Matthew W. Cooper, Jeremy D. Kephart and

Carolyn E. Seifert

*Pacific Northwest National Laboratory, 902 Battelle Boulevard, Richland, WA 99352, USA*

**Abstract**

Pulse shape analysis of germanium gamma-ray spectrometer signals can yield information on the radial position of individual gamma-ray interactions within the germanium crystal. A parametric pulse shape analysis based on calculation of moments of the reconstructed current pulses from a closed-ended coaxial germanium detector is used to preferentially select single-site gamma-ray interactions. The double escape peak events from the 2614.5 keV gamma-ray of $^{208}$Tl are used as a training set to optimize the single-site event selection region in the pulse shape parameter space. A collimated source of 320.1 keV gamma-rays from $^{51}$Cr is used to scan different radial positions of the same semi-coaxial germanium detector. The previously trained single-site selection region is used to preferentially identify the single-site photoelectric absorption events from the 320.1 keV full-energy peak. From the identified events, a comparison of the pulse shape parameter space distributions between different scan positions allows radial interaction location information to be collected. © 2001 Elsevier Science. All rights reserved

## 1. Introduction

Position sensitivity in the bulk of high purity germanium semi-conductor spectrometers enhances the value and utility of high resolution energy measurements of nuclear radiation. The knowledge of the interaction locations are used to track gamma-rays and there-by image gamma-ray sources. Segmentation of semi-conductor detectors is one technique to obtain the desired position information. Pulse shape analysis of the collected signals is another means to extract interaction location information. Reported here is an investigation of radial position reconstruction of single-site gamma-ray events was conducted using a parametric pulse shape analysis technique. The combination of "pencil-beam" gamma-ray collimation and radial reconstruction is intended to selectively study single-site gamma-ray interactions localized in a small region of the bulk of HPGe detector. This article presents the development of a method for extracting radial information from the signal pulses of single-site gamma-ray interactions using a parametric pulse shape analysis technique.

---

[*] Corresponding author. Tel.: +1-509-376-4361; fax: +1-509-376-8002; e-mail: john.orrell@pnl.gov.

## 2. Pulse Shape Analysis

Pulse shape analysis (PSA) as used here means the analysis of the features of the signal pulse due to a gamma-ray interaction in a radiation detector. There are two broad categories of PSA: Library methods and parametric methods. Library based PSA uses a set of reference pulse shapes to match the measured data to the expected pulse shapes for given interaction types or locations in the HPGe crystal [1-4]. Parametric PSA calculates various values (i.e. generically, moments of the signal pulse) that are intended to distinguish different event-type populations by their appearance in different portions of a multi-dimensional space formed from these calculated values [5-6].

A parametric PSA based on the work in Ref. [6] and reported in Ref. [7-8] is used because it has been demonstrated to preferentially select single-site gamma-ray interactions in HPGe detectors. The parametric PSA determines three parameters for each pulse: pulse width, pulse asymmetry, and normalized moment. The pulse width is simply the total time elapsed for the collection of the charge carriers from the gamma-ray interaction in the HPGe detector, based on the calculated current pulse. Based on the pulse width's calculation of the start and end points of the pulse, the mid-time of the pulse is used to calculate a pulse asymmetry value. The asymmetry is defined as the difference between the areas under the front and back halves of the current pulse, divided by the total area. In this asymmetry calculation the area under the current pulse is physically the total charge collected. The final PSA parameter is the normalized moment, which is analogous to a calculation of a mass distribution of a moment of inertia. Together these three parameters form a multi-dimensional parameter shape in which event selection takes place. Figure 1 is an example of the three-dimensional parameter space in which events are located by the values of the three calculated moments discussed above. This data is discussed further below.

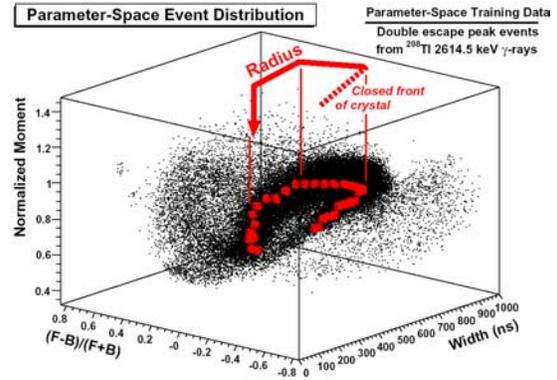

Fig. 1. Pulse shape analysis parameter space filled with double-escape peak events from the 2614.5 keV gamma-ray of $^{208}$Tl. The red highlighted region and schematic arrow show a hypothesized relationship to radial interaction location in the crystal.

## 3. Data Collection

Two data sets were collected. The first exposed the 90 mm diameter HPGe detector to the $^{232}$Th decay chain which includes a 2614.5 keV gamma-ray from $^{208}$Tl. The double escape peak (DEP) events from this gamma-ray were used as a training set for identifying the region of PSA parameter space populated by single-site events, such as double escape events. The second data set was of a collimated source of 320.1 keV gamma-rays from $^{51}$Cr. Simulation results show approximately 15% of events contributing to the 320.1 keV full energy peak interact in the crystal via single-site photoelectric absorption. The result of the training-set acts as a filter to preferentially select events composing the 15% of interest. The collimator was oriented perpendicular to the axis of the closed-ended coaxial HPGe detector. The collimated gamma-ray beam was directed toward the "waist" of the detector where the crystal's electric field lines approximate the electric field lines found in a pure coaxial detector. The collimated gamma-ray beam was scanned across one-half of the detector in 5 mm increments of the impact parameter, ranging from 10 mm – 40 mm. At each impact parameter a different distribution of radii in the crystal are sampled by the attenuated 320.1 keV gamma-ray beam. The second data set provides groups of data with known radial distributions of single-site events. In this way distributions in the PSA parameter space can be correlated with radial distributions, allowing for the possibility of radial interaction position reconstruction of single-site gamma-ray interactions.

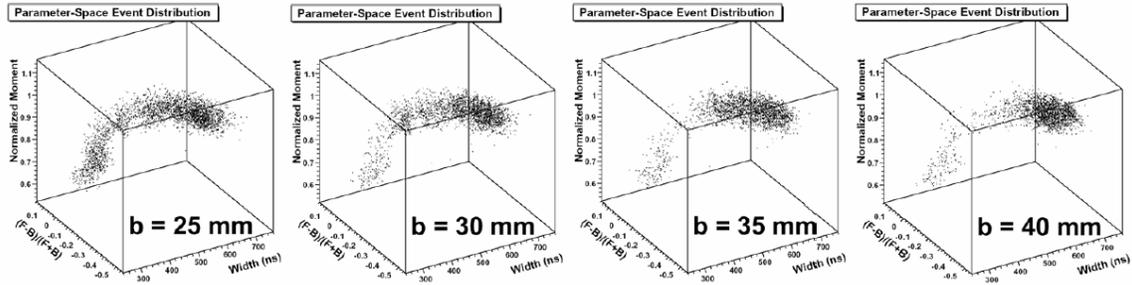

Fig. 2. Collimated 320.1 keV gamma-rays from 51Cr set at an impact parameter *b*. All plots have equal number of events. The redistribution of events as a function of *b* demonstrates the radial information present in the pulse shape analysis parameter space.

**4. Data Analysis**

The analysis begins by calculating the three PSA parameters for events in an energy range from 1500 keV – 1650 keV in the $^{232}$Th data set. The DEP used as the single-site training set is located at 1592.5 keV, very near a full energy peak at 1588.2 keV from $^{228}$Ac, also present in the $^{232}$Th decay chain. With an average resolution (i.e. standard deviation) of 1.27 keV obtained for the data in the training set, these two peaks could only partially be resolved. An energy window from 1591 keV – 1595 was chosen from which to draw examples of single-site events, that is, DEP events. Based on a fit to the peaks and continuum, it is estimated that 67% of the events that fall within this small selected energy range are in fact DEPs. The distribution of events in this 4 keV energy range is shown in Figure 1. The dark, high density regions are expected to be the single-site DEP events. The light, less dense scatter outside this region is believed to be composed of events having multiple interactions in the crystal. From this raw data distribution, the histogram's bins are sorted by density. From this sorted list of bins, the highest density bins are selected as "valid" bins that identify events that are preferentially single-site in nature. This bin selection continues until 67% of the total events in the histogram are accounted for. Once a reduced "valid" region of bins is selected, a path along the contour of the highest density region can be formed, shown as red colored bins in Figure 1. Based on prior experience with this parametric PSA technique, a relationship between positions in the highest density region and the radial position of the single-site interaction in the crystal is hypothesized. This hypothesis is drawn schematically in red in Figure 1. The hypothesis is based on recognition of the nature of the radial drift of charge in the HPGe detectors and the meaning of the three parameters chosen for the PSA. With a path through the high density region found, the work with the training set is complete.

The analysis of the $^{51}$Cr data is simplified by the fact that only the peak location of the 320.1 keV peak needs to be identified in the energy spectrum. That is, for the remainder of this analysis, energy calibration is not a prerequisite. Identifying the 320.1 keV full energy peak for each of the collimated, scanning runs, the three parameters are calculated for events that fall within 50 keV on either side of this peak. The 67% region of "valid" bins determined from the training set is then used as a filter to choose events that we expect to be preferentially of a single-site nature, specially, the approximately 15% of events in the 320.1 keV full-energy peak that are photoelectric absorption events. Upon applying this filter, approximately 30% of the full energy peak remained[1] implying a signal to noise ratio of approximately 1. This is not unexpected since multi-site events do appear in the same region of the PSA parameter space as single-site events. For comparison, without the PSA based filter, the single-site to multi-site ratio would be 15:85 or signal to noise ratio of about 0.18. The four plots in Figure 2 show the distribution of events in the PSA parameter space for different impact parameter settings, $b$, of the collimated source of 320.1 keV gamma-rays from $^{51}$Cr. There are equal numbers of events in each of the four figures. The redistribution of those events as a function of the impact parameter demonstrates there is radial-interaction location-information present. However, similar figures for impact parameters less than 25 mm do not show clear evidence for additional radial information. This may be caused by the fact that there is a radial point of symmetry about which the minimum pulse width and PSA asymmetry parameter are degenerate.

## 5. Conclusions

This work demonstrates parametric pulse shape analysis of signals from a high purity germanium gamma-ray spectrometer can provide information on the radial position of interaction, for single-site gamma-ray interactions. At large radii in the germanium crystal, the analysis presented here shows position resolution of less than 5 mm can be achieved with the parametric pulse shape analysis. For small radii in the germanium crystal, further study of the response of the pulse shape analysis algorithm is needed. In both cases, to quantify the exact radial position resolution a finer-grained procedure of stepping across the impact parameter, $b$, is needed. In many applications, single millimeter or better radial position resolution is desirable and is the desired goal for these methods. Furthermore, study of the net signal to noise ratio after application of the pulse shape analysis filtering is necessary to quantify systematic uncertainties in the radial position resolution due to contamination from multi-site events.

## Acknowledgments

We thank Brian Hyronimus for engineering and assembling the lead collimator and test stand used in this experiment. This work was funded as part of the Radiation Detection Materials Discovery Initiative [9], a Laboratory Directed Research and Development effort at the Pacific Northwest National Laboratory operated by Battelle for the U.S. Department of Energy under Contract DE-AC05-76RL01830.

---

[1] The value is reported as approximate because the actual value varies for each position of the collimator.